\documentclass[epj]{webofc}
\usepackage[varg]{txfonts}   
\wocname{EPJ Web of Conferences}
\woctitle{ICNFP 2016}
  \newcommand{\cB}{{\cal B}}

  \newcommand{\cV}{{\cal V}}

\newcommand{\tr}{{\textrm{tr}}}

\begin{document}
\selectlanguage{english}
\title{Anomaly induced transport in non-anomalous currents~\thanks{Presented by E.~Meg\'{\i}as at the 5th International Conference on New Frontiers in Physics (ICNFP 2016), 6-14 July 2016, Kolymbari, Crete, Greece.}}
%
%

\author{Eugenio Meg\'{\i}as\inst{1,2}\fnsep\thanks{\email{emegias@mppmu.mpg.de}}
}

\institute{Max-Planck-Institut f\"ur Physik (Werner-Heisenberg-Institut), F\"ohringer Ring 6, D-80805, Munich, Germany
\and
          Departamento de F\'{\i}sica Te\'orica, Universidad del Pa\'{\i}s Vasco UPV/EHU, Apartado 644,  48080 Bilbao, Spain
}

\abstract{%
  Quantum anomalies are one of the subtlest properties of relativistic field theories. They give rise to non-dissipative transport coefficients in the hydrodynamic expansion. In particular a magnetic field can induce an anomalous current via the chiral magnetic effect. In this work we explore the possibility that anomalies can induce a chiral magnetic effect in non-anomalous currents as well. This effect is implemented through an explicit breaking of the symmetries. 
}
\maketitle
\section{Introduction}
\label{intro}

The basic ingredients in the hydrodynamic approach are the constitutive relations, which are expressions of the energy-momentum tensor $T^{\mu\nu}$, and the charge currents $J^\mu$, in terms of fluid quantities~\cite{Kovtun:2012rj}. These relations are supplemented with the hydrodynamic equations, which correspond to the conservation laws of the currents. However, in presence of chiral anomalies the currents are no longer conserved, i.e. $\partial_\mu T^{\mu\nu} \ne 0$ and $\partial_\mu J^\mu \ne 0$. This leads to very interesting non-dissipative phenomena that already appear at first order in the hydrodynamic expansion: the {\it chiral magnetic effect}, responsible for the generation of an electric current parallel to a magnetic field~\cite{Fukushima:2008xe}, and the {\it chiral vortical effect}, in which the current is induced by a vortex~\cite{Son:2009tf}. The constitutive relation for the charge currents then~read
\begin{equation}
\langle J^\mu \rangle = n u^\mu + \sigma^\cB B^\mu + \sigma^\cV\omega^\mu + \cdots \,,
\end{equation}
where $n$ is the charge density, $u^\mu$ is the local fluid velocity, $B^\mu = \frac{1}{2}\epsilon^{\mu\nu\rho\lambda} u_\nu F_{\rho\lambda}$ is the magnetic field, with the field strength of the gauge field defined as $F_{\mu\nu} = D_\mu A_\nu - D_\nu A_\mu$, and  $\omega^\mu = \epsilon^{\mu\nu\rho\lambda}u_\nu D_\rho u_\lambda$ is the vorticity vector. The transport coefficients responsible for the chiral magnetic and vortical effects, $\sigma^\cB$ and $\sigma^\cV$, have been studied in a wide variety of methods: these include Kubo formulae~\cite{Amado:2011zx,Landsteiner:2012kd,Chowdhury:2015pba}, diagrammatic methods~\cite{Manes:2012hf}, fluid/gravity correspondence~\cite{Bhattacharyya:2008jc,Erdmenger:2008rm,Banerjee:2008th,Megias:2013joa}, and the partition function formalism~\cite{Banerjee:2012iz,Jensen:2012jy,Jensen:2012jh,Megias:2014mba}. It is already clear from these studies that the axial anomaly~\cite{Kharzeev:2009pj} and the mixed gauge-gravitational anomaly~\cite{Landsteiner:2011cp} are responsible for the previously mentioned non-dissipative effects. 

In this work we are going to focus on the Kubo formalism. The most significant result of anomalies is that they produce equilibrium currents, and the corresponding conductivities are defined via Kubo formulae that involve retarded correlators at zero frequency. The Kubo formula for the chiral magnetic conductivity was derived in~\cite{Kharzeev:2009pj}, and it reads
\begin{equation}
\sigma^\cB = \lim_{p_c\rightarrow 0} \frac{i}{2p_c} \sum_{a,b}\epsilon_{abc}
\langle J^a J^b \rangle|_{\omega=0}  \,. \label{eq:Kubo}
\end{equation}
A similar formula involving the energy-momentum tensor was derived in~\cite{Amado:2011zx} for the chiral vortical conductivity.  In this work we use this formalism to study the chiral magnetic effect of anomalous conductivities. We will study the possibility that, under certain circumstances, this effect might be present in non-anomalous currents as well.

\section{The chiral magnetic and separation effects}
\label{sec:CME}

The chiral magnetic (CME) and chiral separation (CSE) effects are examples of anomalous transport. In the context of heavy ion collisions, a very strong magnetic field produced during a non-central collision induces a parity-odd charge separation which can be modelled by an axial chemical potential, and as a consequence an electric current parallel to the magnetic field is generated, leading to the CME~\cite{Fukushima:2008xe,Kharzeev:2009pj,KerenZur:2010zw}. On the other hand, chirally restored quark matter might give rise to an axial current parallel to a magnetic field, known as CSE~\cite{Newman:2005as}. These effects have been predicted as well in condensed matter systems, see e.g.~\cite{Basar:2013iaa,Landsteiner:2013sja}.

Let us consider a theory of $N$ chiral fermions transforming under a global symmetry group $G$ generated by matrices $(T_A)^f\,{}_g$. The chemical potential for the fermion $\Psi^f$ is given by $\mu^f= \sum_A q_A^f \mu_A$, while the Cartan generator is $H_A = q^f_A \delta^f\,{}_g$ where $q^f_A$ are the charges. The general form of the anomalous induced currents by a magnetic field is
\begin{equation}
\vec{J}^a  = (\sigma^\cB)_{ab}  \vec{B}_b \,, \label{eq:CME}
\end{equation}
where $\vec{B}_b$ is the magnetic field corresponding to symmetry $b$. The 1-loop computation of the chiral magnetic conductivity by using the Kubo formula of Eq.~(\ref{eq:Kubo}) leads to~\cite{Kharzeev:2009pj,Landsteiner:2011cp,Chowdhury:2015pba}
\begin{equation}
(\sigma^\cB)_{ab} = d_{abc} \frac{\mu^c}{4\pi^2}  \,, \qquad  d_{abc} = \frac 1 2 [ \tr( T_a \{ T_b, T_c\} )_R -  \tr( T_a \{ T_b, T_c\} )_L ] \,,
\end{equation}
where $d_{abc}$ is the group theoretic factor related to the axial anomaly, which typically appears in the computation of the anomalous triangle diagram corresponding to three non-abelian gauge fields coupled to a chiral fermion. The subscripts $R$, $L$ stand for the contributions of right-handed and left-handed fermions. Anomalies are responsible for a non-vanishing value of the divergence of the current, that reads in this case~\cite{Kumura:1969wj}
\begin{equation}
D_\mu J_a^\mu= \epsilon^{\mu\nu\rho\lambda} \frac{d_{abc}}{32\pi^2} F^b_{\mu\nu} F^c_{\rho\lambda}    \,.
\end{equation}
Let us particularize Eq.~(\ref{eq:CME}) to the symmetry group $U_V(1) \times U_A(1)$. Then there are vector and axial currents induced by the magnetic field of the vector fields, i.e.
\begin{equation}
\vec{J}_V = \frac{\mu_A}{2\pi^2} \vec{B}_V  \,, \qquad \vec{J}_A = \frac{\mu_V}{2\pi^2} \vec{B}_V  \,, \label{eq:CMECSE}
\end{equation}
which correspond to the CME and CSE respectively. 

The question then arises: is it possible to get a {\it chiral magnetic effect} for a non-anomalous symmetry~$w$? This means to have an induced current in symmetry $w$, i.e.
\begin{equation}
\vec{J}_w \ne 0 \qquad \textrm{with} \qquad    d_{wab} = d_{awb} = d_{abw} = 0 \qquad \forall a,b.
\end{equation}
In the rest of the manuscript we will study the possibility that anomalies can induce transport also in non-anomalous currents.

\section{Holographic model}
\label{sec:holo_model}

The Kubo formula Eq.~(\ref{eq:Kubo}) has been computed in~\cite{Landsteiner:2011iq,Landsteiner:2011tf,Landsteiner:2013aba} at strong coupling within a Einstein-Maxwell model in 5 dim.  In order to account for the anomalous effects, the model is supplemented with Chern-Simons (CS) terms. In this work we will restrict to a pure gauge CS term, which mimics the axial anomaly. The action reads
\begin{equation}
  S = \frac{1}{16\pi G} \int d^5x \sqrt{-g} \bigg[ R + 12  - \frac{1}{4} F_V^2 - \frac{1}{4} F_A^2 - \frac{1}{4} F_W^2 + \frac{\kappa}{3} A \wedge F_V \wedge F_V  \bigg]  + S_{GH} \,, \label{eq:S}
\end{equation}
where $S_{GH}$ is the usual Gibbons-Hawking boundary term. $V_M$ and $A_M$ are vector and axial gauge fields, respectively, and $W_M$ is an extra gauge field associated to the non-anomalous symmetry $w$. The anomalous term mixes the $V_M$ and $A_M$ fields.~\footnote{We use the notation for capital indices $M \in \{r,t,x,y,z\}$, and Greek indices $\mu \in \{t,x,y,z\}$.} A computation of the currents with this model leads~to
\begin{eqnarray}
\tilde{J}_V^\mu &=& \frac{\partial S}{\partial V_\mu}  =  -\lim_{r\to \infty} \frac{\sqrt{-\gamma}}{16\pi G} \left[  F_V^{r \mu} + 6\kappa \epsilon^{\mu\nu\rho\lambda} A_\nu F^V_{\rho\lambda} \right] = J_V^\mu + K^\mu \,, \label{eq:JV} \\
\tilde{J}_A^\mu &=& \frac{\partial S}{\partial A_\mu}  = -\lim_{r\to \infty} \frac{\sqrt{-\gamma}}{16\pi G}  F_A^{r \mu} = J_A^\mu  \,, \\
\tilde{J}_W^\mu &=& \frac{\partial S}{\partial W_\mu}  = -\lim_{r\to \infty} \frac{\sqrt{-\gamma}}{16\pi G} F_W^{r \mu} = J_W^\mu \,.
\end{eqnarray}
$\tilde{J}^\mu$ stands for the holographic {\it consistent} currents, which are not gauge covariant in general. The {\it covariant} version of the currents, denoted by $J^\mu$, is the usual one appearing in the constitutive relations, and it corresponds to the covariant part, dropping the CS current~$K^\mu$. An analysis of the divergence of the covariant currents leads to the {\it "covariant'' anomalies}:
\begin{equation}
D_\mu J_{V}^\mu =  -3\kappa \epsilon^{\mu\nu\rho\lambda} F^A_{\mu\nu} F^V_{\rho\lambda} \,, \qquad D_\mu J_{A}^\mu = -\frac{3}{2}\kappa \epsilon^{\mu\nu\rho\lambda} F^V_{\mu\nu} F^V_{\rho\lambda} \,,  \qquad D_\mu J_{W}^\mu = 0  \,.
\end{equation}
From this result, it is clear that while one expects the existence of anomalous transport effects in $J_{V}^\mu$ and $J_{A}^\mu$, this is not the case for $J_{W}^\mu$. In a holographic computation of the conductivities with this model, one gets the result of Eq.~(\ref{eq:CMECSE}) for $J_V^\mu$ and $J_A^\mu$, but a vanishing value for $J_W^\mu$.

\subsection{Holographic model with symmetry breaking}
\label{sec:sym_break}

In the following we are going to study the possibility that the constitutive relation for $\langle J_W^\mu \rangle$ receives anomalous contributions. Let us extend the model of Eq.~(\ref{eq:S}) with the contribution ($S_{\textrm{tot}} = S + S_\phi$)
\begin{equation}
 S_\phi = \frac{1}{16\pi G} \int d^5x \sqrt{-g} \bigg(  - |D_M \phi|^2 - m^2\phi^2  \bigg) \,, \qquad  D_M \phi = \left[ \partial_M -i(A_M - W_M) \right] \phi  \,, \label{eq:Sphi}
\end{equation}  
where $\phi$ is a scalar field with a tachyonic bulk mass $m^2 = \Delta (\Delta -4)$, and $0 \le \Delta \le 4$. $S_\phi$ produces an explicit breaking of $A$ and $W$ symmetries via the scalar field~$\phi$. From the {\it AdS/CFT dictionary}, the model is the holographic dual of a Conformal Field Theory (CFT) with a deformation
\begin{equation}
{\cal L} = {\cal L}^{CFT} + \lambda \, {\cal O} \,,
\end{equation}
where ${\cal O}$ is an operator dual of the scalar field with $\dim {\cal O} = \Delta$, and $\lambda$ is the source of the operator with $\dim \lambda = 4-\Delta$. The near boundary expansion of~$\phi$ reads
\begin{equation}
\phi(r) = \phi_{4-\Delta} r^{\Delta-4} + \phi_{\Delta} r^{-\Delta} + \cdots \,, \qquad r \to \infty \,, \label{eq:phir}
\end{equation}
where $\phi_{4-\Delta}$ is interpreted as the source $\lambda$, and $\phi_\Delta$ as the condensate $\langle {\cal O}\rangle$. In the following we will choose $\Delta = 3$, so that the bulk mass is $m^2 = -3$. The explicit breaking of symmetries is realized via the boundary term $\lim_{r\to\infty} r \cdot \phi(r) = M$, where we have identified the mass parameter $M$ with the source in Eq.~(\ref{eq:phir}). Our goal is to study the induced currents
\begin{equation}
\vec{J}_V = \sigma_V(M) \vec{B} \,,  \qquad \vec{J}_A = \sigma_A(M) \vec{B} \,, \qquad \vec{J}_W = \sigma_W(M) \vec{B} \,, \label{eq:currents}
\end{equation}
where one expects a dependence of the conductivities in $M$. A non zero value for $\sigma_W(M)$ would signal the existence of a non-anomalous current induced by anomalies.

\subsection{Background equations of motion}
\label{sec:bkg}

 We will work in the probe limit, so that the metric fluctuations are neglected. The equations of motion of the background $(\delta g^{MN}, \delta V^M, \delta A^M, \delta W^M, \delta \phi)$ can be solved by considering the AdS Schwarzschild solution
\begin{equation}
ds^2 = -r^2f(r) dt^2 + \frac{dr^2}{r^2f(r)} + r^2\left( dx^2 + dy^2 + dz^2\right) \,, \qquad f(r) = 1 - \frac{r_h^4}{r^4} \,,
\end{equation}
and the background gauge fields
\begin{equation}
V = V_t(r)dt \,, \qquad A = A_t(r)dt \,, \qquad W = W_t(r)dt  \,. 
\end{equation}
The chemical potentials are computed as $\mu_Y \equiv Y_t(r\to\infty) - Y_t(r_h)$ with $Y = V, A, W$. Then, the fields have the following near boundary expansion
\begin{eqnarray}
&&\lim_{r\to\infty} r \cdot \phi(r) = M \,, \label{eq:nearboundaryphi}  \\
&&\lim_{r\to\infty} V_t(r) = \mu_V + V_t(r_h) \,, \quad \lim_{r\to\infty} A_t(r) = \mu_A + A_t(r_h) \,, \quad \lim_{r\to\infty} W_t(r) = \mu_W + W_t(r_h) \,. \label{eq:nearboundaryVAX}
\end{eqnarray}
It is convenient to define in the following new fields $A_1 = A - W$ and $A_2 = A + W$, so that the covariant derivative writes~$D_M\phi = \left[ \partial_M -i(A_1)_M \right] \phi$. The boundary expansions of~$A_{1,2}$ write as in Eq.~(\ref{eq:nearboundaryVAX}), with chemical potentials $\mu_{1,2} \equiv \mu_A  \mp \mu_W$. Then, the equations of motion of the background read
\begin{eqnarray}
0 &=& V_t^{\prime\prime} + \frac{3}{r} V_t^\prime \,, \qquad 0 \; =  \; (A_2)_t^{\prime\prime} + \frac{3}{r} (A_2)_t^\prime \,, \label{eq:Vt} \\
0 &=& (A_1)_t^{\prime\prime} + \frac{3}{r} (A_1)_t^\prime - \frac{4 \phi^2}{r^2f(r)}(A_1)_t \,, \label{eq:A1t}  \\ 
0 &=& \phi^{\prime\prime} + \left( \frac{5}{r} + \frac{f^\prime}{f} \right) \phi^\prime + \left( \frac{(A_1)_t^2}{r^4f^2} - \frac{m^2}{r^2f} \right) \phi \,. \label{eq:phi} 
\end{eqnarray}
The solutions of $V_t$ and $(A_2)_t$ can be obtained analytically, and the result is\begin{equation}
V_t(r) = c_V - \mu_V \frac{r_h^2}{r^2} \,, \qquad (A_2)_t(r) = c_2 - \mu_2 \frac{r_h^2}{r^2} \,,
\end{equation}
where $c_V$ and $c_2$ are integration constants. The solutions of $A_1$ and $\phi$ can be obtained numerically by solving the coupled system of differential equations, Eqs.~(\ref{eq:A1t})-(\ref{eq:phi}). From these equations one can easily see that regularity of the solution near the horizon demands $(A_1)_t(r_h) = 0$.

\section{Conductivities in presence of symmetry breaking}
\label{sec:Conductivities}

The Kubo formulae for the conductivities appearing in Eq.~(\ref{eq:currents}) are:
 \begin{equation}
\sigma_{Y} = \lim_{p_z\to\infty} \frac{i}{p_z} \langle (J_{Y})_x (J_V)_y \rangle|_{\omega=0}  \,, \qquad \textrm{with} \qquad Y = V, A, W \,. \label{eq:sigmaY}
\end{equation}
These involve retarded correlators which can be computed by using the AdS/CFT dictionary, see e.g.~\cite{Son:2002sd,Herzog:2002pc,Kaminski:2009dh,Landsteiner:2012kd}. We will explain in this section the computation in some details.

\subsection{Fluctuations}
\label{sec:Fluctuations}

Without loss of generality we consider perturbations of momentum $p$ in the $z$-direction at zero frequency. To study the effect of anomalies  it is enough to consider the shear sector, i.e. transverse momentum fluctuations,
\begin{equation}
{\cal K}_\mu = K_\mu(r) + \epsilon k_\mu(r) e^{i p_z z}  \,, \qquad \mu = x,y \,,
\end{equation}
where $K_\mu(r)$ refers to the background solutions computed in Sec.~\ref{sec:bkg} for any of the fields $V$, $A_{1,2}$ (or $V$, $A$ and $W$), and $k_\mu(r)$ are the corresponding fluctuations $v$, $a_{1,2}$ (or $v$, $a$ and $w$).

In studying the fluctuations it is useful to organize the equations of motion according to their helicity under the transverse SO(2) rotation, which is a left-over symmetry. The equations for the fluctuations of the gauge fields are then classified into helicity $\pm 1$, and at order ${\cal O}(\epsilon)$ they read~\footnote{The equation for the scalar field is only affected by these perturbations at order ${\cal O}(\epsilon^2)$.}
\begin{eqnarray}
0 &=& v_{\pm}^{\prime\prime} + \left( \frac{3}{r} + \frac{f^\prime}{f} \right) v_\pm^\prime - \frac{p_z}{3r^4f} \big( 3p_z \pm 4\kappa r \left( (A_1)_t^\prime + (A_2)_t^\prime \right) \big) v_\pm  \mp \kappa \frac{4p_z}{3r^3f} \left( (a_1)_\pm + (a_2)_\pm \right) V_t^\prime   \,,  \label{eq:v} \\
0 &=& (a_1)_{\pm}^{\prime\prime} + \left( \frac{3}{r} + \frac{f^\prime}{f} \right) (a_1)_\pm^\prime - \frac{1}{r^4f} \big( p_z^2 + 4 r^2\phi^2 \big) (a_1)_\pm  \mp \kappa \frac{8p_z}{3r^3f} V_t^\prime v_\pm \,, \label{eq:a1} \\
0 &=& (a_2)_{\pm}^{\prime\prime} + \left( \frac{3}{r} + \frac{f^\prime}{f} \right) (a_2)_\pm^\prime - \frac{p_z^2}{r^4f} (a_2)_\pm  \mp \kappa \frac{8p_z}{3r^3f} V_t^\prime v_\pm   \,, \label{eq:a2}
\end{eqnarray}
where the fields are defined as  $v_\pm = v_x \pm i v_y$ and $(a_{1,2})_\pm = (a_{1,2})_x \pm i (a_{1,2})_y$. In order to obtain the retarded correlation functions we should perform a low momentum expansion of the fluctuation solutions, i.e.
\begin{equation}
v_\pm = v^{(0)}_\pm + p_z v^{(1)}_\pm + \cdots  \,, \qquad (a_{1,2})_\pm = (a_{1,2})^{(0)}_\pm + p_z (a_{1,2})^{(1)}_\pm + \cdots  \,.
\end{equation}
From Eqs.~(\ref{eq:v})-(\ref{eq:a2}) one gets the corresponding equations at order ${\cal O}(p_z^0)$ and  ${\cal O}(p_z)$, which can be solved by imposing the appropriate boundary conditions for the fluctuation fields. In general we should consider regularity of the fields up to the horizon, and the sourceless condition, which means that the fluctuations cannot modify the background fields at the boundary. Then
\begin{equation}
k_\mu(r_h) = \textrm{finite} \,, \qquad k_\mu(r \to \infty ) = 0 \,,  
\end{equation}
for any of the fluctuation fields $k_\mu = v_\pm$ or $(a_{1,2})_{\pm}$. The only fluctuation that can be computed analytically is $(a_2)_{\pm}$, and the solution at order ${\cal O}(p_z)$ reads
\begin{equation}
(a_2)_\pm^{(1)}(r) = \mp c \kappa \frac{4}{3} \frac{\mu_V}{r_h^2} \log\left( 1 + \frac{r_h^2}{r^2} \right) \,, \label{eq:sola21}
\end{equation}
where $c$ is an integration constant. In a near boundary expansion, the fields behave as
\begin{eqnarray}
v_\pm^{(1)} &=& C_{0,v} + \frac{\widetilde{C}_{2,v}}{r^2}\log r + \frac{C_{2,v}}{r^2} + \cdots \,, \\
(a_1)_{\pm}^{(1)} &=& C_{0,a_1} + \frac{\widetilde{C}_{2,a_1}}{r^2}\log r + \frac{C_{2,a_1}}{r^2} + \cdots \,, \label{eq:expa1} \\
(a_2)_{\pm}^{(1)} &=& C_{0,a_2} + \frac{\widetilde{C}_{2,a_2}}{r^2}\log r + \frac{C_{2,a_2}}{r^2} + \cdots \,.
\end{eqnarray}
The sourceless condition for the fluctuations imply $C_{0,v} = C_{0,a_1} = C_{0,a_2}=0$. The other coefficients $\widetilde{C}_2$, $C_2$ can be obtained from a numerical solution of the equations of motion of the fluctuations.

\subsection{Conductivities}
\label{sec:conductivities}

\begin{figure*}[htb]
\begin{tabular}{cc}
\includegraphics[width=6.65cm]{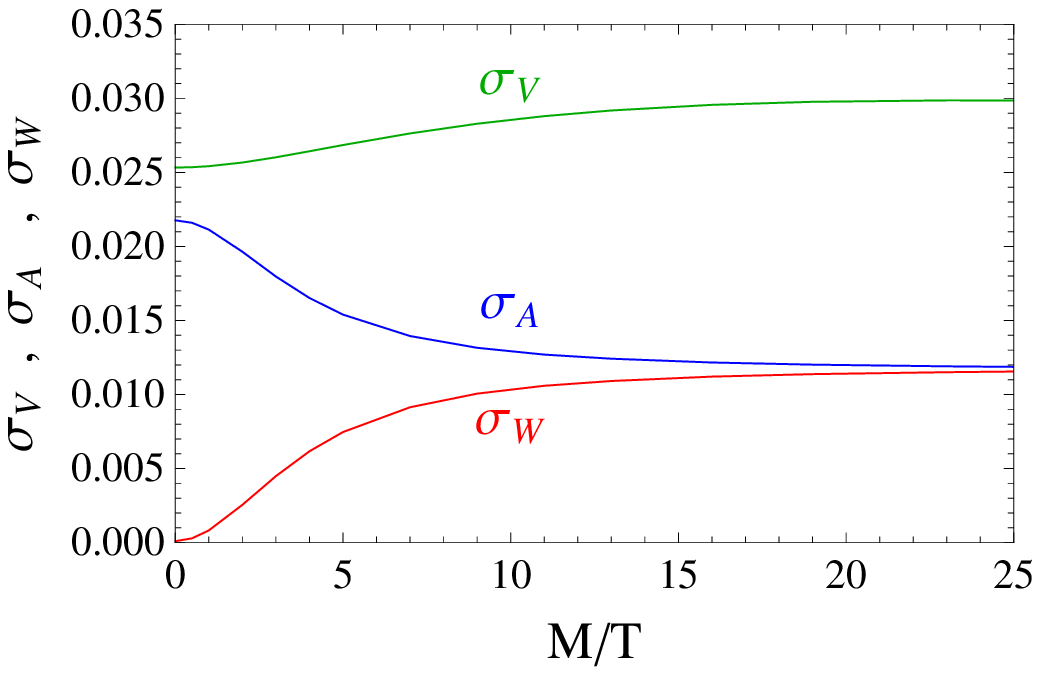} & 
\includegraphics[width=6.65cm]{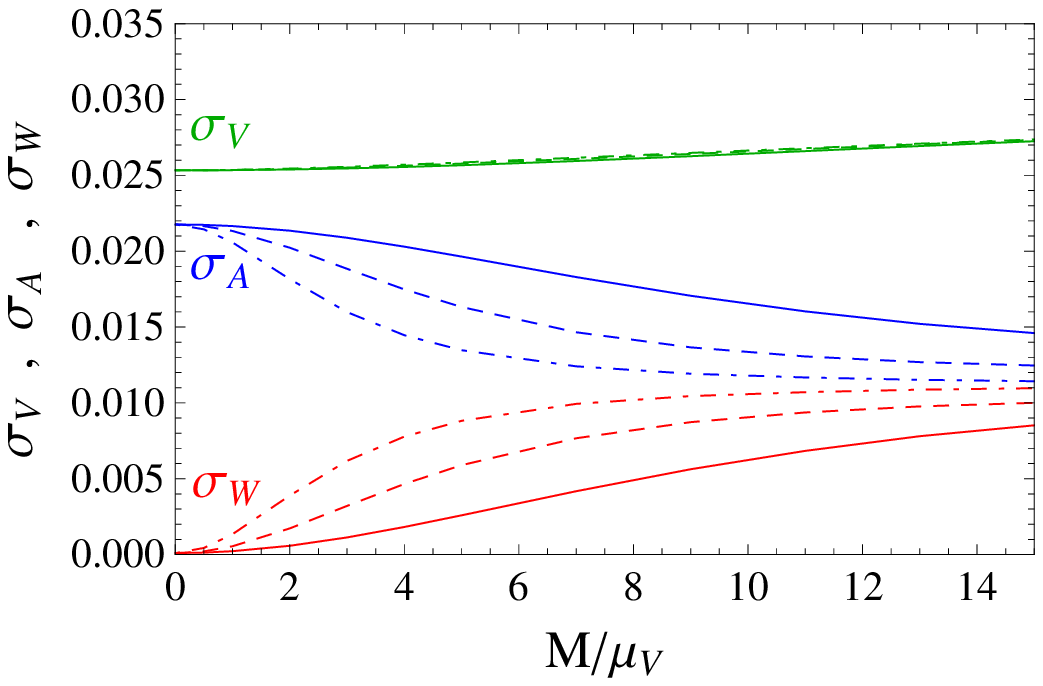}
\end{tabular}
\caption{\it Vector $\sigma_V$, axial $\sigma_A$ and induced $\sigma_W$ conductivities as a function of $M/T$ (left) and $M/\mu_V$ (right). Left panel: we have considered $\mu_V=0.4$, $\mu_A=0.5$, $\mu_W=0.6$ and $T=1$. Right panel: We have considered the same values of the chemical potentials, for temperatures $T=1$ (solid), $T=0.5$ (dashed) and $T=0.3$ (dot dashed).}
\label{fig:conductivities}
\end{figure*}

The correlators of Eq.~(\ref{eq:sigmaY}) are contained in the coefficients $C_{2,v}$, $C_{2,a_1}$ and $C_{2,a_2}$. In particular, we have the following contributions
\begin{equation}
C_{2,v} \!=\!  \lim_{p_z \to 0} \frac{i}{p_z} \! \sum_{Y=V,1,2}  \!\!\! \langle J_V J_Y \rangle|_{\omega=0}   \,,  \quad\!\!\! C_{2,a_1} \!=\! \lim_{p_z \to 0} \frac{i}{p_z} \! \sum_{Y=V,1,2} \!\!\! \langle J_1 J_Y \rangle|_{\omega=0}   \,, \quad\!\!\! C_{2,a_2} \!=\!  \lim_{p_z \to 0} \frac{i}{p_z} \! \sum_{Y=V,1,2} \!\!\! \langle J_2 J_Y \rangle|_{\omega=0}   \,. \label{eq:Cv2}
\end{equation}
Using the relation between the conductivities in the $(A,W)$ and $(A_1,A_2)$ basis
\begin{equation}
\sigma_{1,2} \equiv \lim_{p_z \to 0} \frac{i}{p_z} \langle J_{1,2} J_V \rangle = \sigma_A \mp \sigma_W \,, \label{eq:sigma12}
\end{equation}
one can express them as
\begin{equation}
\sigma_V = \lim_{p_z \to 0} \frac{i}{p_z} \langle J_V J_V \rangle|_{\omega=0}  \,, \qquad \sigma_{A,W} = \lim_{p_z \to 0} \frac{i}{2 p_z}\left( \langle J_2 J_V \rangle \pm \langle J_1 J_V \rangle  \right)|_{\omega=0}    \,, \label{eq:conductivities_result}
\end{equation}
where the sign $\pm$ corresponds to the case $\sigma_A$ and $\sigma_W$ respectively. We show in Fig.~\ref{fig:conductivities} the result for the chiral conductivities of Eq.~(\ref{eq:conductivities_result}) after solving numerically the equations of motion of the fluctuations. We find a non zero value for $\sigma_W(M)$ in presence of explicit symmetry breaking, i.e. $M \ne 0$. In particular, we observe the following properties:
\begin{itemize}
\item $\sigma_W(0)=0\,.$
\item $\sigma_W(\infty) = \sigma_A(\infty) = \frac{1}{2} \sigma_A(0) \,.$
\end{itemize}
The first property is just the expected behavior of the conductivity of the non-anomalous symmetry~$W$ in absence of symmetry breaking. The second one can be understood intuitively in the following way: In the basis $(A_1,A_2)$ the CS term in the action is
\begin{equation}
S_{CS} = \frac{1}{16\pi G} \int d^5x \sqrt{-g} \bigg[ \frac{\kappa}{6} \left( A_1 \wedge F_V \wedge F_V   +  A_2 \wedge F_V \wedge F_V   \right) \bigg]  \,, \label{eq:SCS}
\end{equation}
so that both symmetries $A_1$ and $A_2$ have a CS interaction. The scalar field however breaks only $A_1$. At $M=0$ there will be CSE for both $A_1$ and $A_2$, but for large $M$, $A_1$ will be badly broken and the CSE in the $J_1$ current goes to zero.~\footnote{See~\cite{Jimenez-Alba:2015awa} for a similar effect.} Since $\sigma_1$ vanishes at $M \to \infty$, we find from Eq.~(\ref{eq:sigma12}) the second~property.

\section{Discussion and outlook}
\label{sec:conclusions}

In this work we have studied the role played by the axial anomaly in the hydrodynamics of relativistic fluids in presence of {\it external magnetic fields}. The anomalous conductivities have been computed by using Kubo formulae. The most interesting result is the characterization of a novel phenomenon related to the possibility that, when symmetries are explicitly broken, anomalies can induce transport effects not only in anomalous currents, but also in non-anomalous ones, i.e. those with a vanishing divergence. We have studied this phenomenon at strong coupling in a holographic Einstein-Maxwell model in 5 dim supplemented with a pure gauge Chern-Simons term. The symmetry breaking is introduced through a scalar field which is dual to an operator ${\cal O}$ with $\dim {\cal O} = 3$. 

We plan to extend this study to other anomalous coefficients, like the chiral vortical conductivity. In addition, it would be interesting to check whether the mixed gauge-gravitational anomaly could induce any effect as well in non-anomalous currents. These and other issues will be addressed in detail in a forthcoming publication~\cite{Megias:workinprogress}.

%
%
%

\begin{acknowledgement}
I would like to thank S.D.~Chowdhury, J.R.~David, K.~Jensen, and especially K.~Landsteiner for valuable discussions.  This work has been supported by Plan Nacional de Altas Energ\'{\i}as Spanish MINECO grant FPA2015-64041-C2-1-P, and by the Spanish Consolider Ingenio 2010 Programme CPAN (CSD2007-00042). I thank the Instituto de F\'{\i}sica Te\'orica UAM/CSIC, Madrid, Spain, for their hospitality during the completion of the final stages of this work. The research of E.M. is supported by the European Union under a Marie Curie Intra-European fellowship (FP7-PEOPLE-2013-IEF) with project number PIEF-GA-2013-623006, and by the Universidad del Pa\'{\i}s Vasco UPV/EHU, Bilbao, Spain, as a Visiting Professor.
\end{acknowledgement}


\end{document}